\documentclass[8pt]{article}
\usepackage[utf8]{inputenc}

\usepackage[english]{babel}

\usepackage[a4paper,top=3cm,bottom=2cm,left=3cm,right=3cm,marginparwidth=1.75cm]{geometry}

\usepackage{amsmath,amssymb,amsthm}
\usepackage{algorithm}
\usepackage{algpseudocode}
\usepackage{graphicx}
\usepackage{bbm,enumerate}
\usepackage[colorlinks=true, allcolors=blue]{hyperref}
\usepackage{bbm}
\usepackage{comment}
\usepackage[export]{adjustbox}
\usepackage{subcaption}
\usepackage{authblk}

\theoremstyle{plain}

\title{Probabilistic Record Linkage of Two Gun Violence Data Sets}
\author[1, *]{Iris Horng}
\author[2]{Qishuo Yin}
\author[3]{William Chan}
\author[4]{Jared Murray}
\author[5]{Dylan S. Small}
\affil[1]{Department of Statistics and Data Science, University of Pennsylvania, Philadelphia}
\affil[2]{Department of Operations Research and Financial Engineering, Princeton University, Princeton}
\affil[3]{Department of Economics, University of Pennsylvania, Philadelphia}
\affil[4]{Department of Statistics, University of Texas at Austin, Texas}
\affil[5]{Department of Statistics and Data Science, University of Pennsylvania, Philadelphia}
\affil[*]{Corresponding author: Iris Horng, University of Pennsylvania, Philadelphia, PA, 19104, USA; 6098282389, ihorng@wharton.upenn.edu}

\date{Mar 2 2023}

\begin{document}

\maketitle

\begin{abstract}
\textbf{Objective:} Gun violence is a serious public health problem in the United States. The Gun Violence Archive (GVA) provides detailed geographic information, while the National Violent Death Reporting System (NVDRS) offers demographic, socioeconomic, and narrative data on gun homicides. We developed and tested a method for merging datasets to inform analysis and strategies to reduce gun violence rates in the United States. 

\textbf{Methods:} After preprocessing the data, we used a probabilistic record linkage program to link records from the GVA $(n = 36,245)$ with records from the NVDRS $(n = 30,592)$. We evaluated sensitivity (the false match rate) by using a manual approach.

\textbf{Results:} The linkage returned $27,420$ matches of gun violence incidents from the GVA and NVDRS datasets. Because of restricted details accessible from GVA online records, only $942$ of these matched records could be manually evaluated. Our framework achieved a $90.12\%$ $(849 \text{ of } 942)$ accuracy rate in linking GVA incidents with corresponding NVDRS records.

\textbf{Practice Implications:} Electronic linkage of gun violence data from $2$ sources is feasible and can be used to increase the utility of the datasets.
\end{abstract}

\section{Introduction}
Gun violence is a serious public health problem in the United States. In 2021, a record high of $81\%$ of homicides were committed with a firearm, marking the highest percentage for homicide by firearm in more than 50 years and highlighting the role of firearms in homicides \cite{cdc_2021_facts}. However, this trend is not evenly distributed across all people in the United States. Research shows widening racial disparities. In the United States in 2021, Black males had the highest age-adjusted rate of firearm-related homicides ($52.9$ deaths per $100 000$ standard population) compared with all other females and people in other racial and ethnic groups \cite{cdc_race}. This racial disparity highlights the disproportionate effect of firearm violence on some populations, emphasizing the need to better understand disparities in gun homicide rates to inform public policies.

Recent work has explored these racial and ethnic disparities at the neighborhood level. Using information on gun homicide deaths from the 2014-2018 Gun Violence Archive (GVA) and information on racial composition of neighborhoods from the 2014-2018 US Census tracts, a modeling study by Cheon et al \cite{Cheon} found that, from 2014 to 2018, regardless of socioeconomic status, the rate of gun homicide deaths increased with the proportion of Black residents \cite{Cheon}. The GVA enabled this neighborhood-specific analysis because of its inclusion of address locations and geographic coordinates (longitude and latitude) \cite{NVDRS_methodology}. However, because the GVA does not provide demographic data (age, sex, race) on the people involved in the incidents, Cheon et al \cite{Cheon} considered only the racial composition and average socioeconomic status of neighborhoods and not the race or socioeconomic characteristics of those who lost their lives in gun homicides. Relying only on this neighborhood information limited the scope of questions that could be addressed. For example, it is unclear whether the higher rate of gun homicides in middle-class, majority-Black neighborhoods compared with middle-class, majority-White neighborhoods was because more middle-class Black residents were dying in these neighborhoods or because majority-Black, middle-class neighborhoods were more likely than majority-White, middle-class neighborhoods to be located near poor neighborhoods. In such cases, residents from poor neighborhoods may spend time in nearby middle-class neighborhoods (because of shared institutions such as schools, grocery stores, and government offices) \cite{peterson2010divergent} and experience violence there. 

To address the limitations of the GVA, researchers may turn to the National Violent Death Reporting System (NVDRS), which, unlike the GVA, provides demographic data on those involved in a gun violence incident and in-depth data on the circumstances surrounding the incident, such as the type of firearm used and how the shooting occurred. However, NVDRS alone is insufficient for analyzing the relationship between neighborhood racial composition and gun violence, because its geographic data only extend to the zip-code level, which Gobaud et al \cite{gobaud2022absolute} note is an imperfect proxy for neighborhood analysis. Therefore, combining the strengths of both datasets—GVA’s precise geographic information and NVDRS’s detailed individual and incident-level data—creates opportunities to address previously unanswerable questions about the intersection of individual and neighborhood characteristics in gun violence. The merged dataset enables analysis of how individual-level demographic characteristics (from NVDRS) interact with neighborhood-level firearm homicide rates (from GVA), uncovering relationships that neither dataset could reveal alone.

\section{Materials and Methods}
\subsection{Models and Algorithms}
\begin{figure}[h]
    \centering
    \includegraphics[scale=0.5]{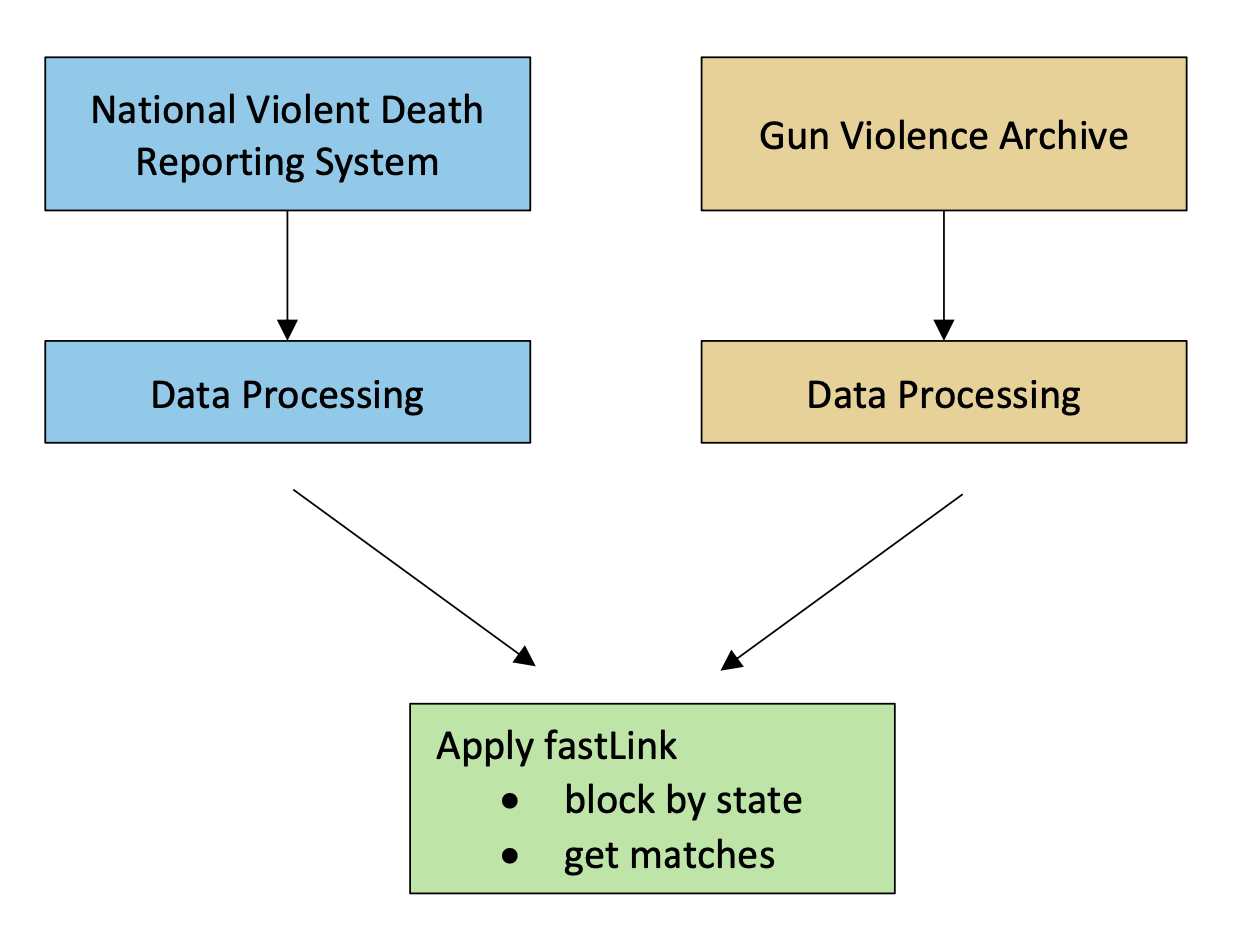}
    \caption{Workflow for record linkage of gun violence data from $2$ datasets, GVA and NVDRS, United States, 2014-2018. The process begins with data processing for the NVDRS and GVA. Next, fastLink is applied, where records are first blocked by state, then probabilistic matching is performed within each state. Finally, matched records are merged from each state to obtain the final linked dataset. Abbreviations: GVA, Gun Violence Archive; NVDRS, National Violent Death Reporting System.}
    \label{fig:workflow}
\end{figure}

We formulated a procedural framework (Figure \ref{fig:workflow}) that can be used to create a linkage between 2 datasets. Each dataset was processed, as described hereinafter. We then applied probabilistic record linkage, specifically the method from Enamorado et al \cite{Enamorado}, to link the GVA and NVDRS datasets, as described hereinafter.

\subsection{Data sources} 
GVA provides yearly records for incidents of gun violence and gun crime and includes data that span all 50 states and the District of Columbia. GVA defines gun violence as any incident involving death or injury or threat with firearms, whether or not intentional, including murder/suicides (where a person kills individuals and then himself), hate crimes, drug involvement, and nearly 120 other possible variables \cite{GVA_methodology}. GVA excludes suicides, except in cases of murder/suicides and officer-involved events. In addition to detailing the date and how many people were injured, GVA data include geographic information at the neighborhood level (down to address). Because GVA contains the longitude and latitude of the incident but not the zip code, we used a geocoding service offered by Texas A\&M University to generate additional zip code and census tract variables \cite{Cheon}. GVA data are available from 2014 onward. At the start of this study, the most recent data available covered the years 2014-2018, so we considered GVA records for those years. 

The NVDRS is a deidentified, multistate, case-level dataset that includes hundreds of unique variables and narratives to describe incidents of all types of violent deaths, including suicides, homicides, unintentional firearm deaths, deaths of undetermined intent, and legal intervention deaths, for all age groups \cite{NVDRS_methodology}. NVDRS data contain demographic characteristics of the people affected and event-specific details for each incident dating to 2002. Available information for each incident includes weapons used (eg, firearm, nonpowder gun, hanging, drowning), cause of death (eg, gunshot wound of head, cardiopulmonary trauma, brain injury), and number injured \cite{NVDRS_methodology}. When we began this study, NVDRS records were available until 2018. To align with the GVA data available at the start of this study, we focused on NVDRS records from 2014 through 2018. These records collectively span all $50$ states, the District of Columbia, and Puerto Rico from 2014 through 2018.

\subsection{Data Processing}
To merge additional details from NVDRS into GVA, we had to refine the NVDRS dataset by processing the data (Figure \ref{fig:workflow}). Compared with GVA, NVDRS captures a broader range of data on violent deaths, many of which fall outside the scope of GVA’s focus on firearm-related incidents. To ensure compatibility between the 2 datasets, we first filtered the NVDRS data (ie, restricted the data to include incidents relevant to our analysis) by excluding single and multiple suicides, because GVA does not report suicides \cite{GVA_methodology}. We retained deaths of undetermined intent in NVDRS (defined as deaths with some evidence of intent but without enough evidence to definitively classify the death as purposeful) because, unlike suicides, these cases may still involve firearm-related violence that aligns with GVA’s scope. Next, we filtered the NVDRS dataset to retain cases in which either the weapon used was a firearm (the reported weapon used must have been either a firearm or a nonpowder gun) or the cause of death involved a firearm (ie, the reported cause of death contained 1 of the following words: gun, firearm, gunshot, rifle). Doing so ensured that any potentially relevant firearm-related NVDRS incidents in GVA’s scope remain for potential matching, allowing for an accurate integration of NVDRS data into the GVA.

To facilitate the merging process, we made additional adjustments to both datasets. To enable numerical comparison of incident dates, we created a new variable, daysSinceStart, which represented the number of days since January 1, 2014, that the incident occurred. Additionally, we standardized the labels of variables of interest, such as incident city and zip code, to ensure consistency. Although both datasets include data for all 50 states and the District of Columbia, NVDRS lacks sufficient coverage for certain states during 2014-2018. To account for data availability, we filtered both datasets to include only records from 40 states and the District of Columbia that reported sufficient information to NVDRS during 2014-2018, as identified in CDC Surveillance Summaries \cite{states2014,states2015,states2016,states2017,states2018} for each year. These jurisdictions included Alaska, Colorado, Georgia, Kentucky, Maryland, Massachusetts, Michigan, New Jersey, New Mexico, North Carolina, Ohio, Oklahoma, Oregon, Rhode Island, South Carolina, Utah, Virginia, and Wisconsin (2014-2017); Arizona, Connecticut, Hawaii, Kansas, Maine, Minnesota, New Hampshire, New York, and Vermont (2015-2018); Illinois, Indiana, Iowa, Pennsylvania, and Washington State (2016-2018); California, Delaware, District of Columbia, Nevada, and West Virginia, (2017-2018); and Alabama, Louisiana, Missouri, and Nebraska (2018).

The resulting cleaned GVA dataset contained $21$ variables for each of the $36\,245$ incidents of gun violence that occurred from 2014 through 2018. The resulting cleaned NVDRS dataset contained 328 variables to describe each of the $30\,592$ incidents of gun violence that occurred from 2014 through 2018. The University of Pennsylvania Institutional Review Board (IRB) determined this study was exempt because it used publicly available data without personal identifiers (IRB $\#856387$).

\subsection{From Deterministic to Probabilistic Matching}
When merging $2$ datasets, the ideal scenario would be to have a unique identifier linking records. For example, if both datasets included the victim’s name for each record, then researchers could easily link the datasets using those unique names. However, such information is typically not available because of privacy concerns, and in datasets such as the NVDRS, victims’ names are never provided.

Instead, deterministic matching can often be used. In this case, rules are used to determine matches, but this method is prone to measurement error \cite{Enamorado}. For example, $2$ records could be declared a match if both incidents occurred in the same city. However, even a small discrepancy, such as a misspelling in the city name, would prevent a match, even if the incidents were truly the same.

To address these limitations, we used the implementation of a canonical probabilistic record linkage by Enamorado et al \cite{Enamorado}, which accounts for such discrepancies by probabilistically merging datasets based on similarities in variables. This approach uses a model to estimate the probability of $2$ records being a match; those pairs whose probability of being a match exceed a chosen threshold are then chosen as matches. In the following section, we describe the key components and framework of the model.

\subsection{Canonical Probabilistic Record Linkage Model}

The canonical model of probabilistic record linkage, originally proposed by Fellegi and Sunter \cite{fellegi1969theory}, merges $2$ datasets that lack unique identifiers by assessing the similarity between records across multiple fields. The model uses agreement patterns across variables to estimate match probabilities (ie, the probability that $2$ records are a match) \cite{Enamorado}. For example, the variables used to determine the probability of $2$ records being a match might include the incident date, incident city, and incident zip code.

For measurement errors or missing data, the model is still able to assign a match probability to a pair of records by using Bayes’ rule \cite{sadinle2017bayesian, sariyar2012missing}, which allows the model to update the prior probability of a match (before actually seeing the extent of agreement between fields for a pair of records) by incorporating the likelihood of observing the given agreement pattern whether or not the pair is indeed a match \cite{Enamorado}. This process results in a posterior probability that the pair is a match, given the observed data. In this way, even if errors in measurement occur or data are missing, the model can estimate the likelihood that $2$ fields for a pair of records agree.

The match probabilities, with values ranging from $0$ to $1$, that the model estimates indicate how likely it is that pairs of records represent the same incident. The higher the match probability, the more confident the model is that the pair is a true match. This probability helps to determine whether to treat the pair of records as a match in the data-merging process. Specifically, a threshold is typically chosen. This means that records with a match probability above the threshold are declared matches, and records with a match probability below the threshold are declared not a match.

In practice, we used \emph{fastLink}, an algorithm developed by Enamorado et al \cite{Enamorado} that implements a canonical model of probabilistic record linkage to merge our datasets. We first declared a threshold for the merging procedure. If a pair of records had match probability above the given threshold, then we declared the pair to be a true match. Enamorado et al \cite{Enamorado} suggest declaring matches using a default threshold value of $0.85$. Using a threshold allows us to control the false discovery rate and the false negative rate to control the correctness of matched pairs.

\subsection{Electronic Linkage Implementation} 
We aimed to identify records in both datasets that belonged to the same victim of an incident of gun violence by using fastLink version $0.6$ (Imai), an open-source software package in R that implements a canonical model of probabilistic record linkage.

\begin{figure}[h]
    \centering
    \includegraphics[scale=0.3]{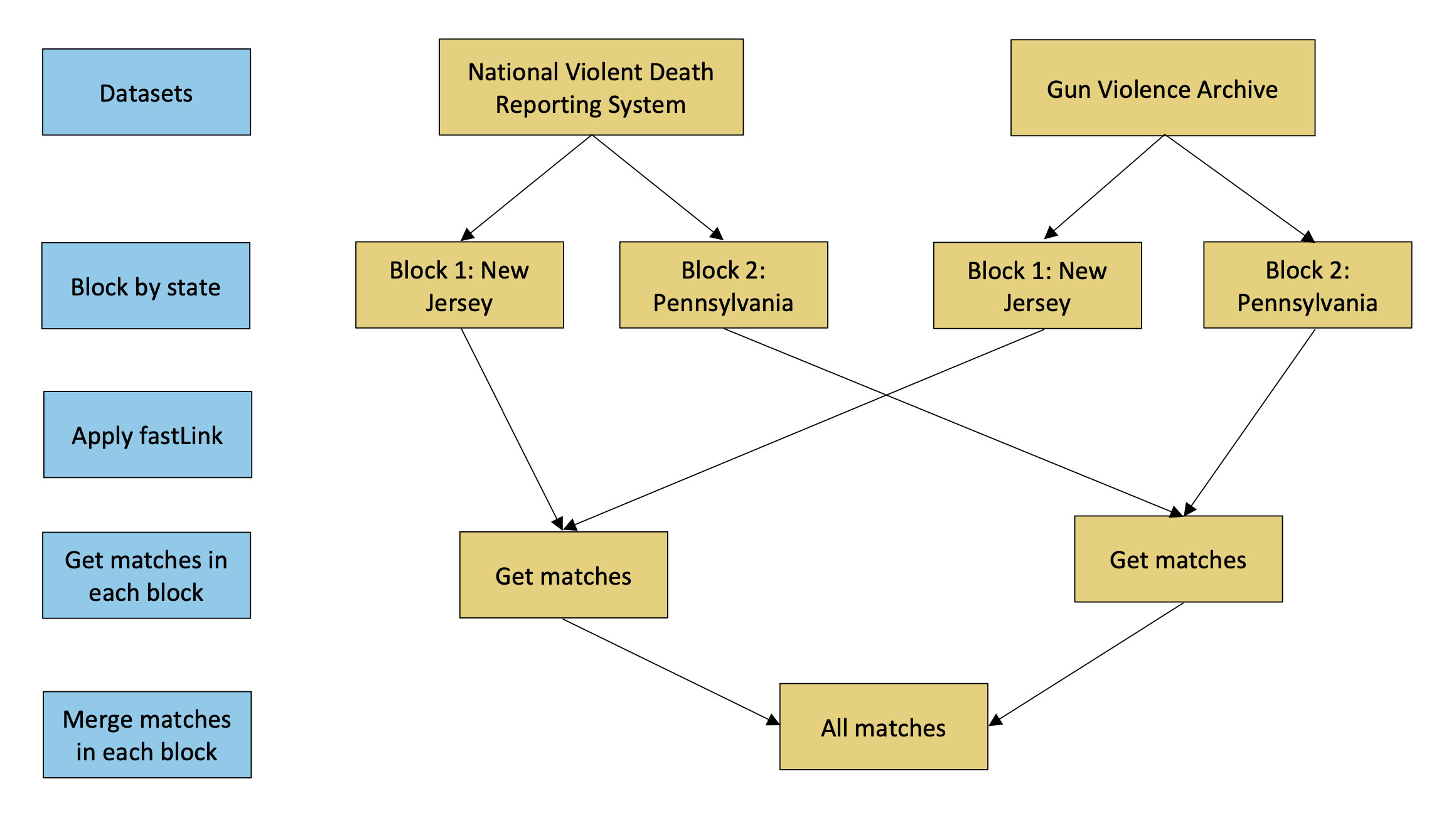}
    \caption{Example of applying fastLink for firearm homicide data from GVA and NVDRS, New Jersey and Pennsylvania, 2014-2018. Assuming GVA and NVDRS contain firearm homicide incidents from $2$ states, New Jersey and Pennsylvania, to perform record linkage, incidents are first blocked by state and the data are partitioned so that each block contains only records from a single state. For each state, records are taken from GVA and NVDRS and fastLink is applied to identify matching records. This process is repeated separately for each state, getting the matches in each block. Finally, all matches from each state block are merged to obtain the final linked dataset. Abbreviations: GVA, Gun Violence Archive; NVDRS, National Violent Death Reporting System.}
    \label{fig:fastlink-procedure}
\end{figure}

Using fastLink’s internal functions, we blocked by state for computational efficiency, grouping the data so that each block contains only records from a single state. Within each block, we merged data by using the following $4$ variables of interest: \textit{InjuryCity} (the city where the incident occurred), \textit{daysSinceStart} (the number of days since January 1, 2014, that the incident occurred), \textit{NumKilled} (number killed in the incident), and \textit{InjuryZip} (the zip code where the incident occurred). We calculated the similarity of the \textit{InjuryCity} using the Jaro-Winkler similarity score (range, 0 to 1), with higher values indicating greater agreement \cite{Winkler}. The remaining $3$ variables used numeric matching. Additionally, we used fastLink’s deduplication procedure to incorporate a limited one-to-one matching constraint in our record linkage process (Figure \ref{fig:fastlink-procedure}) \cite{Jared}. Our code can be found at \href{https://github.com/irishorng/RecordLinkage_GunViolenceIncidents}{GitHub Link}.\cite{Horng_RecordLinkage_GunViolenceIncidents_2024}

\section{Results}
\begin{figure}[h]
    \centering
    \includegraphics[scale=0.3]{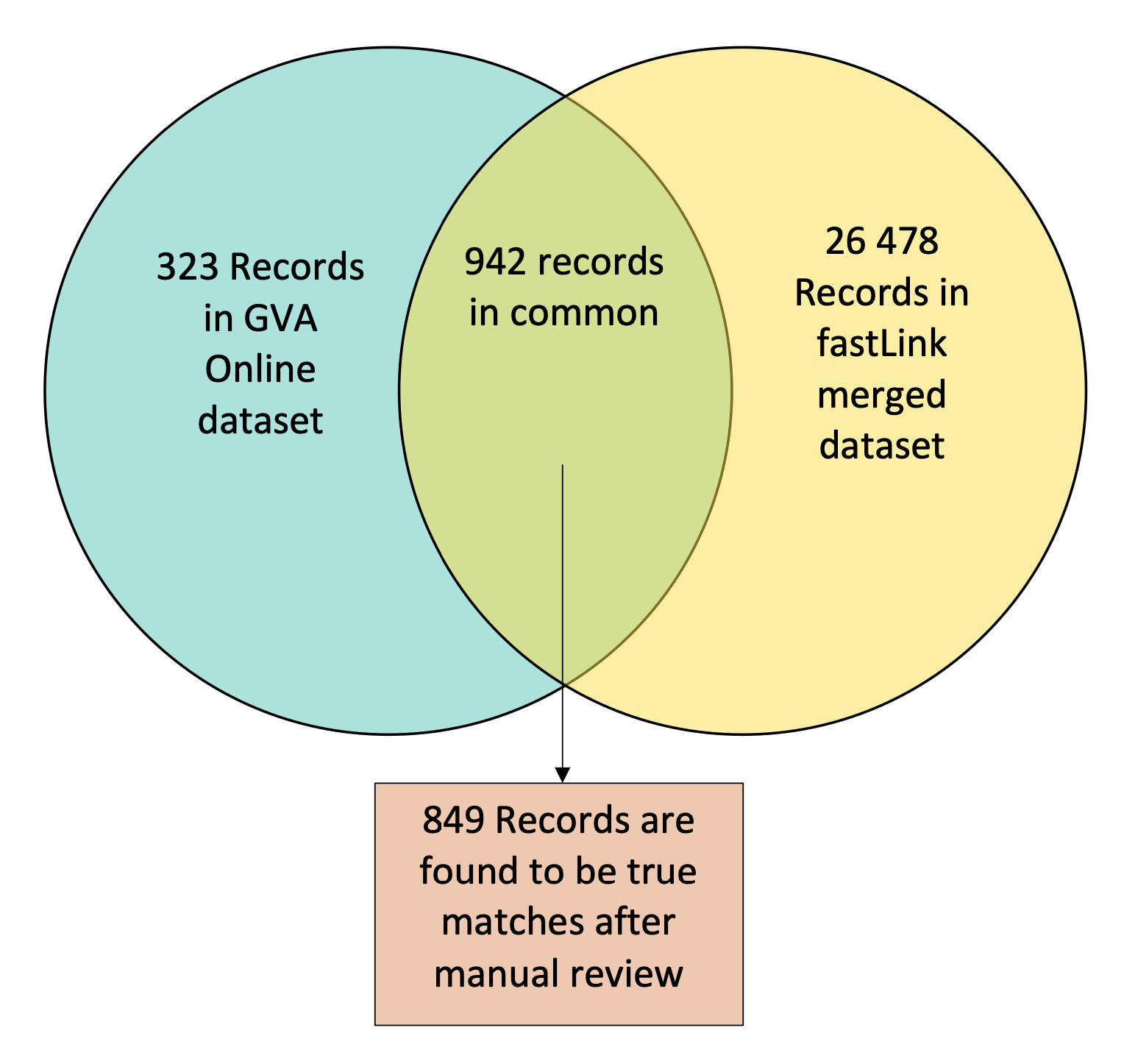}
    \caption{Estimating sensitivity of record linkage for gun violence data from GVA and NVDRS, United States, 2014-2018. The fastLink merged dataset has $27\,420$ records for gun violence incidents. Separately, $1\,265$ records are in the GVA Online dataset, which was obtained from GVA standard reports available online. Comparing these $2$ datasets, $942$ records of gun violence homicides (from $696$ unique incident IDs) were found to be in common. Of these $942$ records, the NVDRS narrative was manually compared with descriptions from the GVA Online dataset and $849$ records were true matches. Abbreviations: GVA, Gun Violence Archive; ID, identification; NVDRS, National Violent Death Reporting System.}
    \label{fig:relationship}   
\end{figure}

Using the GVA dataset of $36\,245$ observations and the NVDRS dataset of $30\,592$ observations and applying a fastLink threshold of 0.85, the process returned $27\,420$ matches of gun violence incidents ($26\,478 + 942$; Figure \ref{fig:relationship}). For each incident, fastLink merged the GVA variables and NVDRS variables. In the following section, we explore the accuracy of the returned matches (Figure \ref{fig:relationship}). 

\subsection{Estimating Sensitivity}  
We calculated the false match rate by manually confirming matches declared by the fastLink method. The observations defined $3$ disjoint sets based on our criteria: record pairs classified as a match, record pairs classified as a nonmatch, and record pairs classified as undetermined (when information was insufficient). The criteria involved a qualitative overview of similarities in the victim and perpetrator age and sex as well as the incident characteristics (ie, circumstances surrounding the death).

To manually confirm matches declared by the fastLink method, we compared the records, which already had NVDRS narratives, with GVA standard reports \cite{GVA_standard_reports} available on their website. GVA publicly provides only selected types of gun violence incidents through its online standard reports, and because we were considering only gun violence deaths, we collected the following standard reports from GVA’s website: children killed, teens killed, officer-involved shootings, school shootings, and mass shootings—all years, mass shootings in 2014, mass shootings in 2015, mass shootings in 2016, mass shootings in 2017, and mass shootings in 2018. We combined these reports into a dataset called GVA Online. Because we previously filtered NVDRS and GVA to include data from only $40$ states and the District of Columbia that reported information to NVDRS from 2014 through 2018, we applied the same filtering to the GVA Online dataset to ensure consistency between the datasets. In total, the GVA Online dataset consisted of $13$ variables describing this selection of incidents and $1265$ observations in total ($323 + 942$; Figure \ref{fig:relationship}).

We noted a total of $1235$ Incident IDs in common between the GVA Online dataset and our original GVA dataset, indicating limited available data on GVA’s website. We found $696$ Incident IDs in common between the GVA Online dataset and our fastLink merged dataset. Because it is possible to have multiple records for a single Incident ID, corresponding to multiple deaths from a single gun violence incident, we found a total of $942$ records in common between the combined GVA standard reports and our fastLink merged dataset.

For this subset of observations returned by fastLink and found in the GVA Online dataset, we compared descriptions from the GVA Online dataset with the NVDRS narrative and classified each record as a match, nonmatch, or undetermined match. From the $942$ records, we classified $849$ records ($90.1\%$) as true matches, $72$ records ($7.6\%$) as nonmatches, and $21$ ($2.2\%$) as undetermined matches.

\subsection{Linkage Feasibility}
Not including data cleaning, electronic linkage with fastLink took about $2$ minutes. Manual review of the $942$ records to assess the accuracy of the electronic linkage took several days.

\section{Discussion}
We implemented and assessed an approach for evaluating the record linkage process of gun violence data and found the linkage process to be feasible and sensitive.

We demonstrated that GVA and NVDRS datasets can be accurately linked. The resulting merged dataset provides many event-specific details while pinpointing the exact neighborhoods in which the incident occurred. This level of detail provides new opportunities for public health interventions tailored to both community- and individual-level risk factors, filling a gap left by previous studies relying on single datasets. While this study used the default and recommended threshold parameter of $0.85$ to identify matched records, future research could explore alternative thresholds to balance false discovery rates and false negative rates according to specific research needs or dataset characteristics.

Our linkage code is publicly available on Github \cite{Horng_RecordLinkage_GunViolenceIncidents_2024}. Prospective studies can easily apply our user-friendly code to create a linkage between larger datasets or different subsets from GVA and NVDRS to conduct analysis on gun homicides from a wider range of years or among certain states, for example. The readability of our code also contributes to its expandability. As more data become available, our code could be extended to implement updates or changes, such as additional variables or even data from other sources.

We note that both GVA and NVDRS may contain coding inaccuracies \cite{GVA_methodology, NVDRS_methodology}. For this study, we used probabilistic matching, which allows records to be linked even if they differ in fields of interest, such as victim counts, based on the likelihood that they represent the same event. While manual adjustments could improve the accuracy of the linkage, we aim to develop a method that minimizes the need for such interventions.

Additionally, GVA’s website offers limited standard reports, which restricted our GVA Online dataset. Consequently, only a small subset of fastLink matches could be manually reviewed, and this process may be prone to human error, potentially affecting the sensitivity findings.

\section{Practice Implications}
We showed that the linkage of neighborhood-specific GVA data with event-detailed NVDRS data is feasible and accurate. In particular, fastLink’s \cite{Enamorado} probabilistic record linkage implementation enhanced the accuracy and efficiency of matching records by incorporating a probabilistic model into the matching procedure, providing researchers with additional flexibility to adjust the model parameters, such as thresholds, according to their needs. While this study used only $2$ datasets, the linkage method can be extended to integrate multiple datasets, providing a more detailed analysis. This approach can be used as the basis for future studies investigating factors influencing gun homicides and monitoring patterns over time, potentially generating insights and suggestions for reducing the risk of gun violence. Moreover, this linkage methodology can be applied to other research areas that involve multiple datasets with overlapping traits. By combining data sources with varying strengths, researchers can gain a deeper understanding of complex social issues, advancing the analysis of trends and outcomes in fields beyond gun violence, such as public health and crime research.

\section{Declarations of Conflicting Interests}
The authors declared no potential conflicts of interest with respect to the research, authorship, and/or publication of this article.

\section{Funding}
The authors received no financial support for the research, authorship, and/or publication of this article.

\bibliographystyle{plain}
\bibliography{reference}

\end{document}